\documentclass[twocolumn]{aastex631}
\usepackage{float}
\usepackage{natbib}
\usepackage{enumitem}
\usepackage{color, soul}
\setstcolor{red}
\submitjournal{ApJ}

\shorttitle{Kodaikanal hand drawn archive and application of clustering techniques}

\shortauthors{Priyadarshi et al.}

\begin{document}

\title{Detection of Solar Filaments using Suncharts from Kodaikanal Solar Observatory Archive Employing a Clustering Approach}

\correspondingauthor{Dipankar Banerjee}
\email{dipu@aries.res.in}

\author[0000-0003-2476-1536]{Aditya Priyadarshi}
\affiliation{Arybhatta Research Institute of Observational Sciences, Nainital, India-263001
}
\affiliation{Indian Institute of Astrophysics, 
Bengaluru, India-560034
}

\author[0000-0002-8163-3322]{Manjunath Hegde}
\affiliation{Arybhatta Research Institute of Observational Sciences, Nainital, India-263001
}
\affiliation{Indian Institute of Astrophysics,
Bengaluru, India-560034
}

\author[0000-0003-3191-4625]{Bibhuti Kumar Jha}
\affiliation{Arybhatta Research Institute of Observational Sciences, Nainital, India-263001
}
\affiliation{Indian Institute of Astrophysics,
Bengaluru, India-560034
}

\author[0000-0002-5014-7022]{Subhamoy Chatterjee}
\affiliation{Southwest Research Institute, 1050 Walnut Street \#300, Boulder, CO 80302, USA1}

\author[0000-0002-7762-5629]{Sudip Mandal}
\affiliation{Max Planck Institute for Solar System Research,  Justus-von-Liebig-Weg 3, 37077 Göttingen,
Germany}

\author[0000-0003-4943-4662]{Mayukh Chowdhury}
\affiliation{Amity University, Noida, UP}

\author[0000-0003-4653-6823]{Dipankar Banerjee}
\affiliation{Arybhatta Research Institute of Observational Sciences, Nainital, India-263001
}
\affiliation{Indian Institute of Astrophysics, 
Bengaluru, India-560034
}
\affiliation{Center of Excellence in Space Sciences India, IISER Kolkata, Mohanpur 741246, West Bengal, India}

\begin{abstract}
With over 100 years of solar observations, the Kodaikanal Solar Observatory (KoSO) is a one-of-a-kind solar data repository in the world. Among its many data catalogs, the `suncharts' at KoSO are of particular interest. These Suncharts (1904-2020) are coloured drawings of different solar features, such as sunspots, plages, filaments, and prominences, made on papers with a Stonyhurst latitude-longitude grid etched on them. In this paper, we analyze this unique data by first digitizing each suncharts using an industry-standard scanner and saving those digital images in high-resolution `.tif' format. We then examine the Cycle~19 and Cycle~20 data (two of the strongest cycles of the last century) with the aim of detecting filaments. To this end, we employed `k-means clustering' method, and obtained different filament parameters such as position, tilt angle, length, and area. Our results show that filament length (and area) increases with latitude and the pole-ward migration is clearly dominated by a particular tilt sign. Lastly, we cross-verified our findings with results from KoSO digitized photographic plate database for the overlapping time period and obtained a good agreement between them. This work, acting as a proof-of-the-concept, will kick-start new efforts to effectively use the entire hand-drawn series of multi-feature, full-disk solar data and enable researchers to extract new sciences, such as the generation of pseudo magnetograms for the last 100 years.
\end{abstract}

\section{Introduction} \label{intro}

Solar features such as sunspots, filaments and plages play a crucial role in our understanding of solar magnetism and associated variability. Soon after the discovery of the telescope in early 1600, various observatories as well as few individual observers across the world started documenting some of these solar features, primarily through drawings \citep{Arlt2008, Usoskin2009, SenthamizhPavai2015, Arlt2020, Carrasco2020}. These drawings capture the past behaviour of the Sun and help us to expand the observational baseline in conjunction with more recently made observations that are available through photographic plates and CCD sensors. The hand-drawn solar charts come from a variety of sources across the globe, such as the Royal Observatory of Belgium (ROB), the Mount-Wilson Observatory (MWO), the Specola Solare Ticinese (SST) in Switzerland, and the National Astronomical Observatory of Japan (NAOJ), to name a few. Among these, ROB hosts the most comprehensive sunspot drawing series, spanning from 1940 to 2011. Other archives such as from Meudon, McIntosh archive and Kislovodsk solar station, have full-disk or synoptic maps with multiple hand-drawn solar features for several solar cycles and are available in digital format. Table~\ref{old_stat} provides a comprehensive list of these archives with their respective observation periods. Unique among all of these databases, KoSO has been systematically observing the Sun since 1904 and, most importantly, in three different wavelengths: white light, Ca II K (393.37 nm), and H-$\alpha$ (656.3 nm). These observations are recorded in photographic plates/films\footnote{More information on KoSO plates/films are available at \url{ https://kso.iiap.res.in/new/data}} and simultaneously in suncharts. These suncharts are one-of-a-kind data as they combine full-disc sketches of various solar features such as sunspots, plages, and filaments/prominences for each day of observation beginning in 1904.

\begin{deluxetable*}{cccccc}
\tablecaption{Solar drawings available from different observatories across the globe\label{old_stat}}
\tablewidth{0pt}
\tablehead{
\colhead{Observatory} & \colhead{Solar Features on the Drawing} & \colhead{Duration} & \colhead{Projection} & \colhead{Condition} & \colhead{Digitization Status} \\
}
\startdata
Gustav Sporer & Sunspot & 1861-1894 & -- & -- & 1861-1894 \\
MWO & Sunspot & 1913-2017 & Partial disk & Good & 1913-2017 \\
ROB & Sunspot & 1940-2011 & Full disk & Excellent & 1940-2011 \\
SST & Sunspot & 1981-2017 & Full disk & Fair & 1981-2017 \\
KoSO & Sunspot, Filaments, Plages, Prominences  & 1904-2020 & Full disk & Excellent & 1954-1976 \\
Meudon & Sunspot, Filaments, PILs, Plages, Prominences  & 1919- & Synoptic Maps & Excellent & 1919- \\
McIntosh archive & Sunspot, PILs, Filaments, Plages, Prominences  & 1967- & Synoptic Maps & Excellent & 1967- \\
Kislovodsk solar station & Sunspot, PILs, Filaments, Plages, Prominences  & 1979-2021 & Full disk and Synoptic Maps & Excellent & 1979-2021 \\
\enddata
\end{deluxetable*}

In this paper, we present the detection of solar filaments from KoSO suncharts. In this context, filaments are elongated dark structures against the bright solar disk observed best in H-~$\alpha$ and \mbox{He\,\sc{II}} wavelengths. It is well established that the location of the filaments shows a good correlation with the regions of large-scale magnetic fields in the Sun \citep{Tlatov2016}. Furthermore, filaments are aligned with magnetic neutral lines and hence, are ideal candidates to study the large-scale concentrations of weaker magnetic fields on the solar disk \citep{McIntosh1972, Low1982, Makarov1983}. Full disc magnetograms are available only after the 1970s, hence, several attempts have been made to generate/recreate the magnetic field, called pseudo-magnetograms, for the previous cycles (before 1970) using indirect proxies of the magnetic field, such as Ca II K observations \citep{Pevtsov2016, Mordvinov2020, Gyungin2020, Theodosios2019d}. Since Ca II K intensity only provides the strength of the magnetic field, hence, by combining the observations of solar filaments, which trace magnetic neutral lines, polarities of magnetic field can be identified while reconstructing the pseudo magnetogram \citep{Mordvinov2020}.

Previous attempts of detecting filaments can be classified into two broad categories: (i) use of different image processing techniques in automatic/semi-automatic way \citep{Gao2002, Fuller2005, Yuan2011, Hao2013, Chatterjee2017, Tlatova2017, Mazumder2021} and (ii) use of machine learning techniques \citep{Zhu2019}. Given that the solar images in most historical archives suffer from inherent image artefacts (such as dust marks, scratches etc.), feature detection via traditional image processing techniques tends to produce erroneous results. In this work, we implement the k-means clustering algorithm to automatically extract solar filaments from KoSO suncharts between 1954 and 1976 (23 years). This paper also describes the novel features of KoSO suncharts and the process of digitizing them.

\section{Data} \label{data}
\begin{figure*}[!ht]
        \centering
        \includegraphics[width=\textwidth]{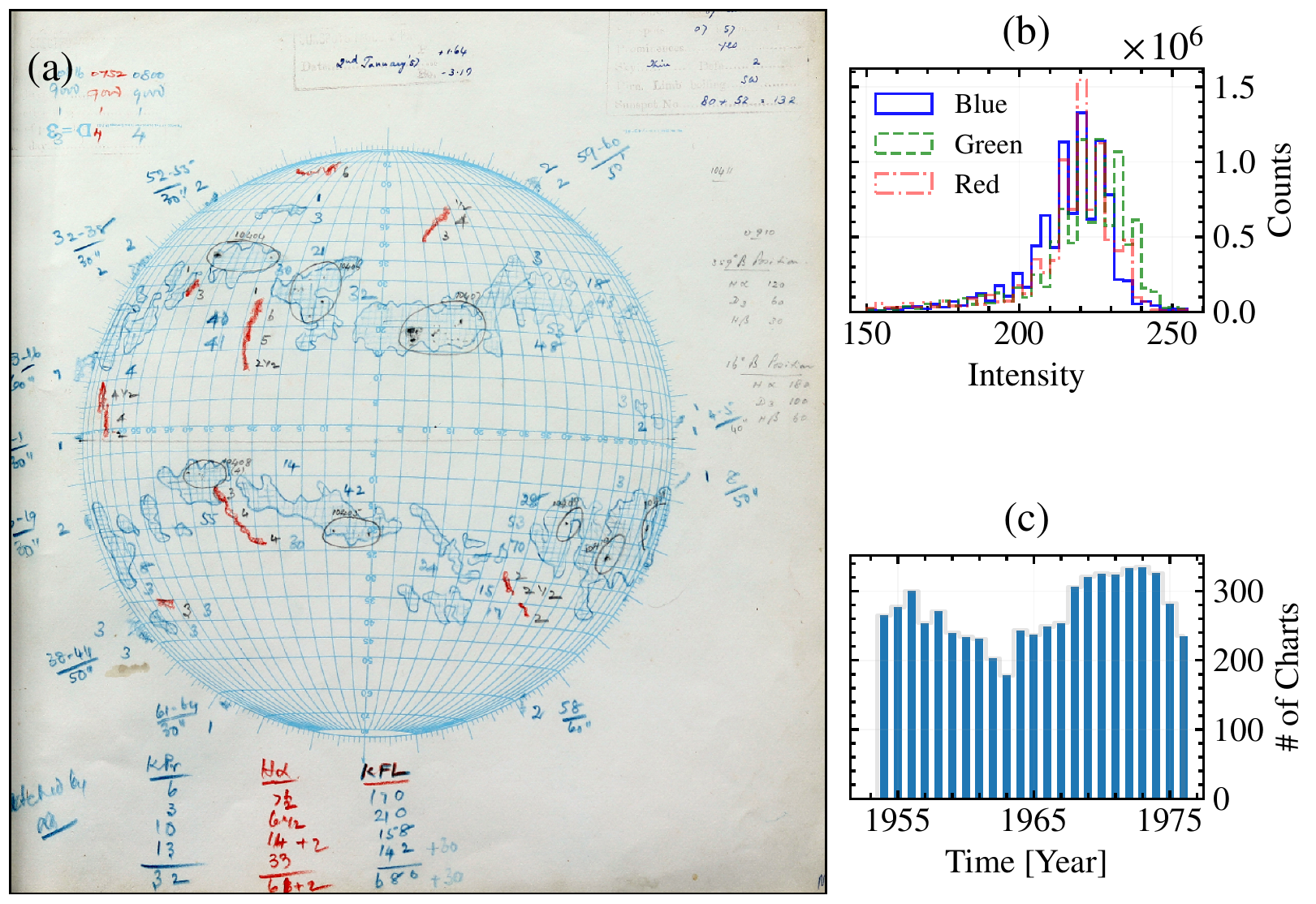}
        \caption{a) A representative image of hand-drawn sunchart made at Kodaikanal Solar Observatory on 2nd January 1957. Sunspots are highlighted in black; different shades denote the umbra and penumbra. The plages are denoted by sky-blue outlines, whereas the filaments are denoted by red. b) The RGB histogram representation for the digitized sunchart. c) Yearly count of digitized KoSO Suncharts.
}
\label{fig1}
    \end{figure*}

In this work, we use KoSO suncharts covering 23 years between 1954 and 1976. Below we outline some of the novel features of these suncharts: Each sunchart (see Figure~\ref{fig1}a) contains several solar features, such as sunspots (from white-light observations), filaments (H-alpha observations), plages and prominences (Ca II K observations) on it. Moreover, all suncharts also have a Stonyhurst latitude and longitude grid (of 5 degrees in size) etched on them. To draw these features, observers used the KoSO photographic plate/film image(s) of that day as reference. As seen in Figure~\ref{fig1}a, every solar feature in these suncharts has a specific colour. For example, sunspots are drawn in black, filaments are in red, and plages and prominences are in blue.
Furthermore, information such as the position of the solar north (P-angle) and the heliographic latitude of the centre of the solar disk (B-angle), along with the time of observation, are written on the top of every chart. Lastly, if there were no observations available from KoSO on a particular day, the observers used data from Meudon and Mt. Wilson Observatory to populate that sunchart. In that case, such features were then drawn with a different colour (e.g., filaments were drawn in green as opposed to their usual red colour) to make it easier to spot such observations. Further information regarding these suncharts and their features can be found in \citet{Ravindra2020}.

\subsection{Digitization of suncharts}
Even though KoSO suncharts cover a period of over 100 years, as a first step, we only digitize 23 years data between 1954 and 1976. The reason behind choosing this period is that it covers cycle-19 and cycle-20, which are the strongest and one of the weakest activity cycles in the last century. Here is a brief account of the digitization procedure. A Canon EOS 800D\footnote{The detail specification of the camera can be found  \url{https://www.canon-europe.com/cameras/eos-800d/specifications/}.} camera with 22.3~mm $\times$ 14.9~mm CMOS sensor is used for digitizing suncharts, and the digitized images are stored in the ``.tiff" (Tagged Image File) format. These digitized images have a bit depth of 24 bits (8 bits $\times$ three channels) with sizes of approximately 3500 $\times$ 3500. Figure~ \ref{fig1}b shows the intensity distributions of a digitized image (in all three channels -Red, Green and Blue), which gives the idea about the sensitivity of the camera sensor for all three colours. The RGB histogram confirms that it can distinguish between different colour spaces.

In Figure~ \ref{fig1}c, we present the statistics of suncharts that have been digitised till now. Furthermore, the figure also highlights the data gap, primarily due to inclement weather at the site, resulting in no observations. Lastly, we mention here that while preparing Figure~\ref{fig1}c as well as the analysis that is to follow, we do not include those suncharts in which the H-$\alpha$ observations were taken from observatories other than KoSO (as mentioned earlier, in those suncharts filaments are sketched with different colour instead of the usual red).

\section{Method}\label{method}
\subsection{Identification of Solar Disk}
\label{ss:disk_id}
\begin{figure*}[!ht]
\centering
\includegraphics[width=0.8\textwidth,clip=]{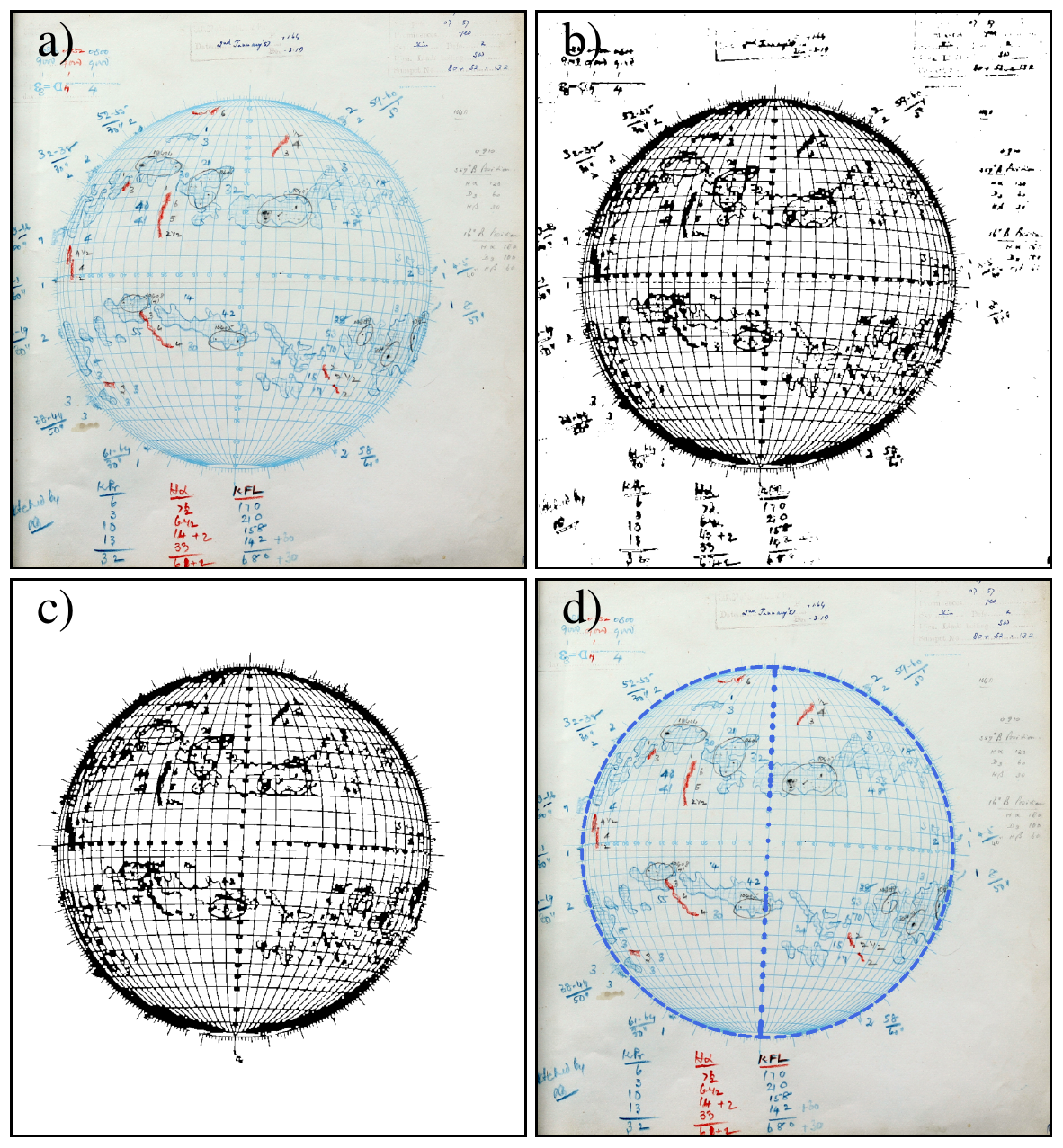}
\caption{(a-d) Numerous steps to detect the solar disc. a) A sample drawing from 2nd January 1957 (same as Figure~\ref{fig1}). b) Output image after applying Canny edge detection and using the morph-close operation, c) shows the largest selected contour, representing the grid from suncharts. Lastly, d) shows the detection of a polar line by Hough transform and detected circle over-plotted over the suncharts grid.}
\label{fig2}
\end{figure*}

The first step towards filament detection is to identify the solar limb in each of these images. To this end, we detect the nearly vertical line in the grid by using the linear Hough transform method \citep{Hough1962}. The radius and centre of the disk are calculated as half of the length of the detected line and its bisection point, respectively. The presence of multiple latitudinal and longitudinal grids, in addition to various overlying markings and features as indicated in Figure  \ref{fig1}a, makes this process of solar limb detection to be a non-trivial one. As a result, before applying the linear Hough transform, we must `clean' the image to remove any undesired features or artefacts, and the steps that we adopted for that are as follows:

\begin{enumerate}[label=\arabic*.]
    \item  First, we process the original image (Figure~\ref{fig2}a) using the Canny edge detection algorithm\footnote{Detail of this function is available at \url{https://www.harrisgeospatial.com/docs/CANNY.html}.}, with a lower and upper threshold of 0.8 and 0.9, respectively. Next, to get rid of those little fragmented structures that the Canny operator often returns, we employ a morph closing operation with a square kernel of the size of 10~pixels. This step produces an image as shown in Figure~\ref{fig2}b.

    \item  Since our interest is in extracting the grid as well as the vertical line, we select the largest connected region in Figure~\ref{fig2}b, that is the grid. The inverted image after this step is shown in Figure~\ref{fig2}c.
    
    \item  Before proceeding to the next step of line identification, we must eliminate the thin extended portions at both ends of the vertical line, which otherwise will lead to overestimating the length. We use a horizontal erosion function with a kernel size of 30 pixels to remove such tentacles and then a horizontal morph close operation with a kernel size of 250 pixels to reconnect the regions. The image is now ready for the linear Hough transform to apply, and in doing so, we determine the line's length, bisection point, and also the slope. These values correspond to the disk's diameter, centre, and orientation in the sunchart, respectively (Figure~\ref{fig2}d). 
\end{enumerate}

\begin{figure*}[!ht]
\centering
\includegraphics[width=\textwidth,clip=]{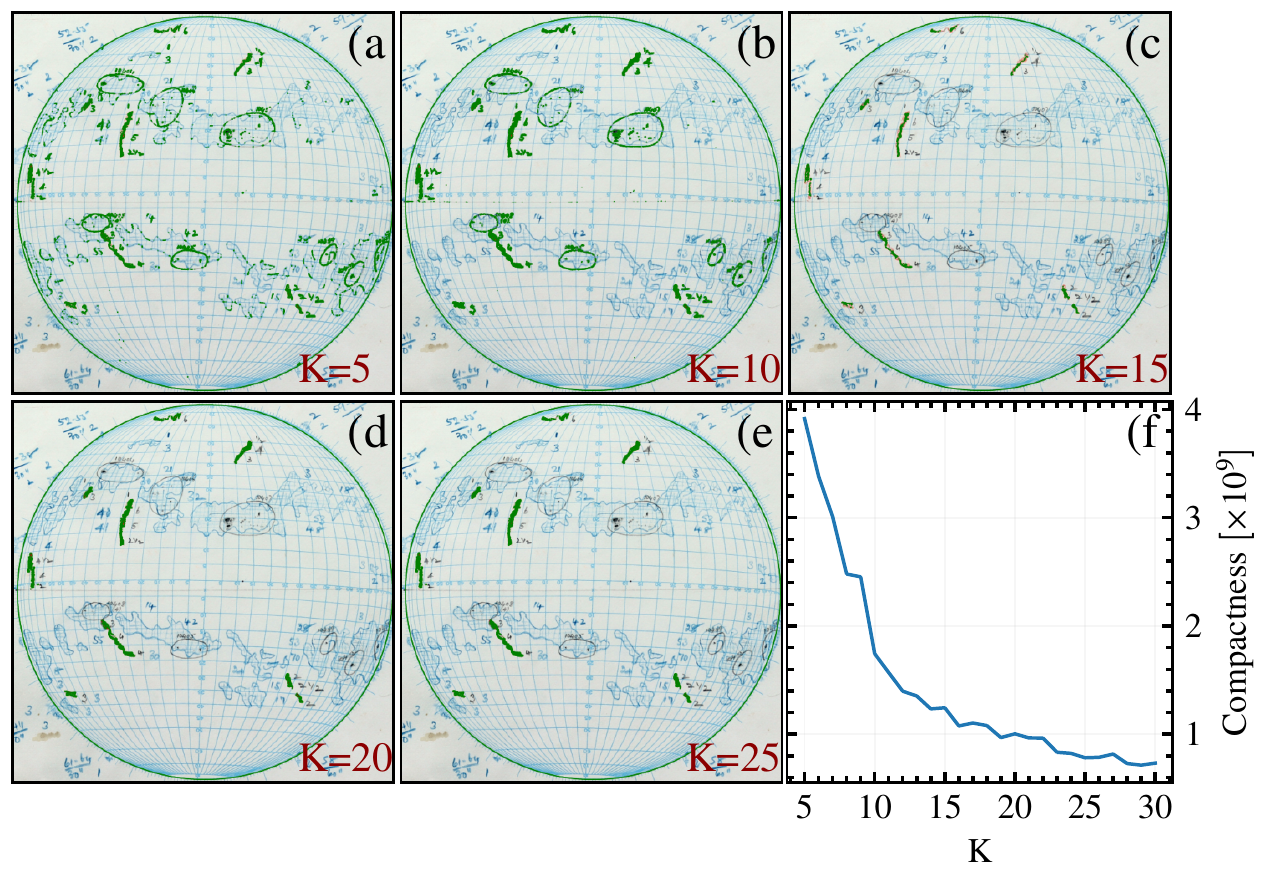}
\caption{ The effect of a number of clusters (K) for K-means clustering in filament detection. a) and b) show the detection of filaments along with other solar features for K = 5 and K = 10, respectively. c) shows filament detection with underestimation of the whole filament for K = 15. d) and e) depict accurate detection of filaments for K = 20, 25 with hardly any difference in visual appearance. f) shows the quantitative estimate of detected filament localization (compactness) as a function of K.
}
\label{fig3}
\end{figure*}

Lastly, although the aforementioned technique works for most suncharts, there are few cases (4\%) where the artefacts are so dominant that our automated limb detection technique does not work and thus, we process those images manually.

\subsection{Identification of Solar Filaments}

We first isolate the disc using the limb information as outlined in Section~\ref{ss:disk_id}. Thus, at this stage, our input image looks similar to the one shown in Figure~\ref{fig2}c. In order to detect the filaments automatically, we employ a clustering algorithm known as K-means clustering \citep{Lloyd1982, MacQueen1967}. The success of this approach lies in identifying the optimal value of K, which describes the number of clusters present in the data. The steps for obtaining this K value are as follows:

\begin{enumerate}[label=\arabic*.]

    \item We first considered all K values between 5 and 30 and visually monitored the output in each run (Figure~\ref{fig3}a\,--\,e). Based on these visual inspections on 100 randomly selected suncharts, we find that K=20 provides the best results. To cross-check this conclusion, we further calculate a quantity known as compactness (a measure of the sum of squared distances between each point and their related centres) and plot it against K as shown in Figure~\ref{fig3}f. We find that the compactness curve initially decreases monotonically with increasing K and then starts to flatten out past K=20. Therefore, it confirms our initial finding of K=20 being the optimal K value. Although the compactness decreases ever so slightly for higher K values (for example, K=30), the computation time increases significantly without much improvement in the final output. Hence, for efficient and effective filament detection, we set K to 20.

    \item The next step is to calculate the mean  RGB values ($\mu_{{\rm R}}$, $\mu_{{\rm G}}$, $\mu_{{\rm B}}$) of the same set of images (as selected in Step~1) for the cluster that best represents the filaments. In addition to the mean, we calculate the mean absolute deviation ($\sigma_{{\rm R}}$, $\sigma_{{\rm G}}$, $\sigma_{{\rm B}}$) in each of the three channels, which offers a range of three RGB values. These ranges are: R=$190$ $\pm$ $20$ and B = $115$ $\pm$ $10$. We see that the G value does not fluctuate from image to image, which is understandable given that filaments are highlighted in red, so we do not use that channel in our case.
    
    \item We apply this range to the R ($\mu_{{\rm R}} \pm \sigma_{{\rm R}}$) and B ($\mu_{{\rm B}} \pm \sigma_{{\rm B}}$) channels and create a binary mask with the same resolution as the cropped sunchart. We denote the filament regions in the mask as 1 and the rest as 0. An inverted (for better visualisation) binary image along with over-plotted contours on sunchart is included in Appendix~\ref{app1}.
    
\end{enumerate}

Upon running the above-mentioned procedure on the KoSO data between 1954-1976 (that include 6594 suncharts), we detected a total of 66722 filaments on them.

\section{Result} \label{result}

\subsection{{Statistical Properties: Time-Latitude diagram}}

\begin{figure*}[!ht]
\centering
\includegraphics[width=0.99\textwidth,clip=]{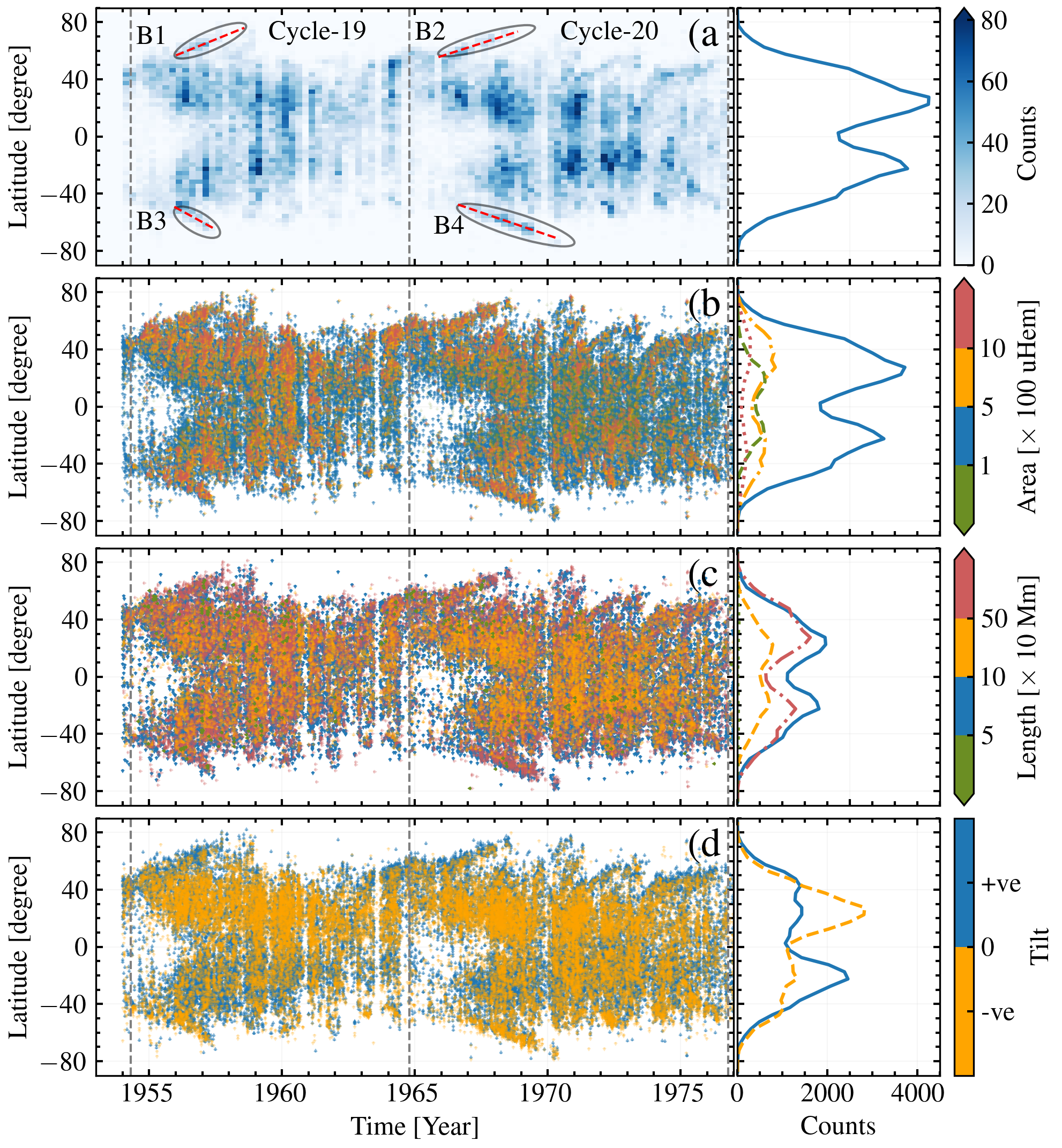}
\caption{a) Distribution of the number of filaments with time and latitudes; here, color represents the number. In the same panel, the indication of polar rushes is marked using ellipses. b), c), and d) show the time-latitude distribution, colored according to the area, length, and sign of tilt of individual filaments. The right panel in each case shows the latitudinal distribution of corresponding parameters over the entire period in consideration.}
\label{fig4}
\end{figure*}

 Different panels of Figure~\ref{fig4} depict the temporal evolution of filament latitudes across the two solar cycles we studied here (Cycles~19 and Cycle~20). In Figure~\ref{fig4}a, we show the well-known 'Butterfly diagram' but through a 2D histogram. The histogram is produced over the time (bin-size $=0.25$ years, i.e., 3 months) and latitude (bin size $=3\degr$) bins. The strength of the colour represents the number of filaments identified in that particular time-latitude bin. Through this plot, we find that although filaments are seem to appear across the entire latitude band, their distribution practically shows a strong cycle dependence. For example, clear signatures of equator-ward and pole-ward migration of filaments are seen during the beginning the cycles. Furthermore, the equator-ward branches are observed to be spread over a larger band of latitude as compared to the sunspots \citep{Sudip2017, Jha2022}. These findings are consistent with ones found previously with digitized H-${\alpha}$ and Ca K photographic plates of KoSO \citep{Chatterjee2017,chatterjee2020}.

Figure~\ref{fig4}b adds further information to the time-latitude diagram of filaments by grouping their de-projected areas (in $\mu$Hem) into different bins and depicting them in different colours. The area ranges considered are: Area $>$ 1000 $\mu$Hem, 500 $\mu$Hem $<$ Area $<$ 1000 $\mu$Hem , 100 $\mu$Hem $<$ Area $<$ 500 $\mu$Hem, and Area $<$ 100 $\mu$Hem.  We find that the filaments with smaller areas are restricted to lower latitudes ($<\pm$35), wherein the bigger ones appear at the higher end. In fact, the pole-ward branches are mostly dominated by such large area filaments. These findings are further confirmed through right panel of Figure \ref{fig4}b, in which we plot the latitudinal distributions of filaments in various area groups collapsing the temporal information.

Lastly, Figure~\ref{fig4}c presents a time-latitude distribution in which we grouped the filaments according to their lengths. Different length ranges that we considered are: length $>$ 500 Mm, 100 Mm $<$ length $<$ 500 Mm, 10 Mm $<$ length $<$ 100 Mm, and length $<$ 10 Mm. To determine the length of a filament in the first place, we start by retrieving the coordinates of a filament's border pixels. Following that, the individual pixel-to-pixel distance (in Mm) are calculated using the respective pixels' latitude and longitude over the spherical surface, and their sum yields the filament perimeter\footnote{We have used the $map\_2points.pro$ function in IDL library (\url{https://www.l3harrisgeospatial.com/docs/map_2points.html)} to calculate inter-pixel distances.}. Filament length is measured as half of this perimeter \citep{Mazumder2018}. Like Figure~\ref{fig4}a, the pole-ward branches in Figure~\ref{fig4}c are also dominated by the longer filaments, wherein the shorter ones are restricted to low latitudes.

\subsection{Tilts of Filaments}
Filaments are typically found along the magnetic polarity inversion lines, and hence, their tilts can be used as a proxy of active region (AR) tilts. Moreover, quantification of AR tilt is essential from the point of view of solar dynamo theory in which this tilt plays a significant role in converting the toroidal field into a poloidal field \citep{Babcock1961, Leighton1964, Arnab2003, Charbonneau2020}. We figure out how much each filament is tilted by using the least-square fit on the spine of each filament. We used the chi-square minimization method to fit a straight line, and then we found the tilt with respect to the sun's equator. Figure~\ref{fig4}d shows the time-latitude distribution of detected filaments colour-coded according to their tilt. Filaments whose spines are oriented counter-clockwise w.r.t. the equator are considered to have positive tilts (highlighted in blue), and the ones which are aligned clockwise are assigned negative tilts (highlighted in orange). From the figure, it is evident that negative tilts dominate the northern hemisphere, whereas positively tilted filaments dominate the southern hemisphere, consistent with the findings of \citet{Mazumder2021} and \citet{Tlatov2016}. This behaviour can also be seen quantitatively in right panel of Figure~\ref{fig4}d.

\subsection{Polar Rush}
All the butterfly diagrams in Figure \ref{fig4} show a common feature which is: during the early phase of a cycle, filaments at higher latitudes (in each hemisphere) migrate towards the pole, which is known as the `polar rush' \citep{Ananthakrishnan1952}. In Figure \ref{fig4}a, we highlight these polar branches through ellipses (B1, B2, B3 and B4). Data points within these ellipses were then retrieved, and a linear fit was applied to estimate the drift rate. Table~ \ref{table_prush} outlines the calculated drift rates for each cycle. Our findings indicate that the northern polar filaments' migration starts before the southern ones' migration. Furthermore, we see a clear North-South asymmetry in migration rates. Furthermore, the drift rates that we obtained from the suncharts match closely with the ones presented in \citet{Xu2021} for Cycle~19 and Cycle~20 (Table~\ref{table_prush}).

\begin{deluxetable}{cccc}[h]
\tablecaption{Drift rate obtained by polar rush fittings from the butterfly diagram.
\label{table_prush}}
\tablewidth{0pt}
\tablehead{
\colhead{Index} & \colhead{($^{\circ}$/year)} & \colhead{($^{\circ}$/CR)  } & \colhead{($^{\circ}$/CR) $^*$}}

\startdata
Cycle 19 N (B1) &  7.49  &  0.55 & 0.51\\ 
Cycle 19 S (B3) &  10.18 &  0.76 & 0.88\\
Cycle 20 N (B2) &  5.99  &  0.44 & 0.29\\
Cycle 20 S (B4) &  6.31  &  0.47 & 0.39\\
\enddata
\tablecomments{ * {\citet{Xu2021}}. The respective branch names are denoted in the Figure \ref{fig4}a. Drift rates are mentioned with their respective units. }

\end{deluxetable}

\subsection{Comparisons with H-${\alpha}$ Carrington maps}
\begin{figure}[h]
\centering
\includegraphics[width=\columnwidth,clip=]{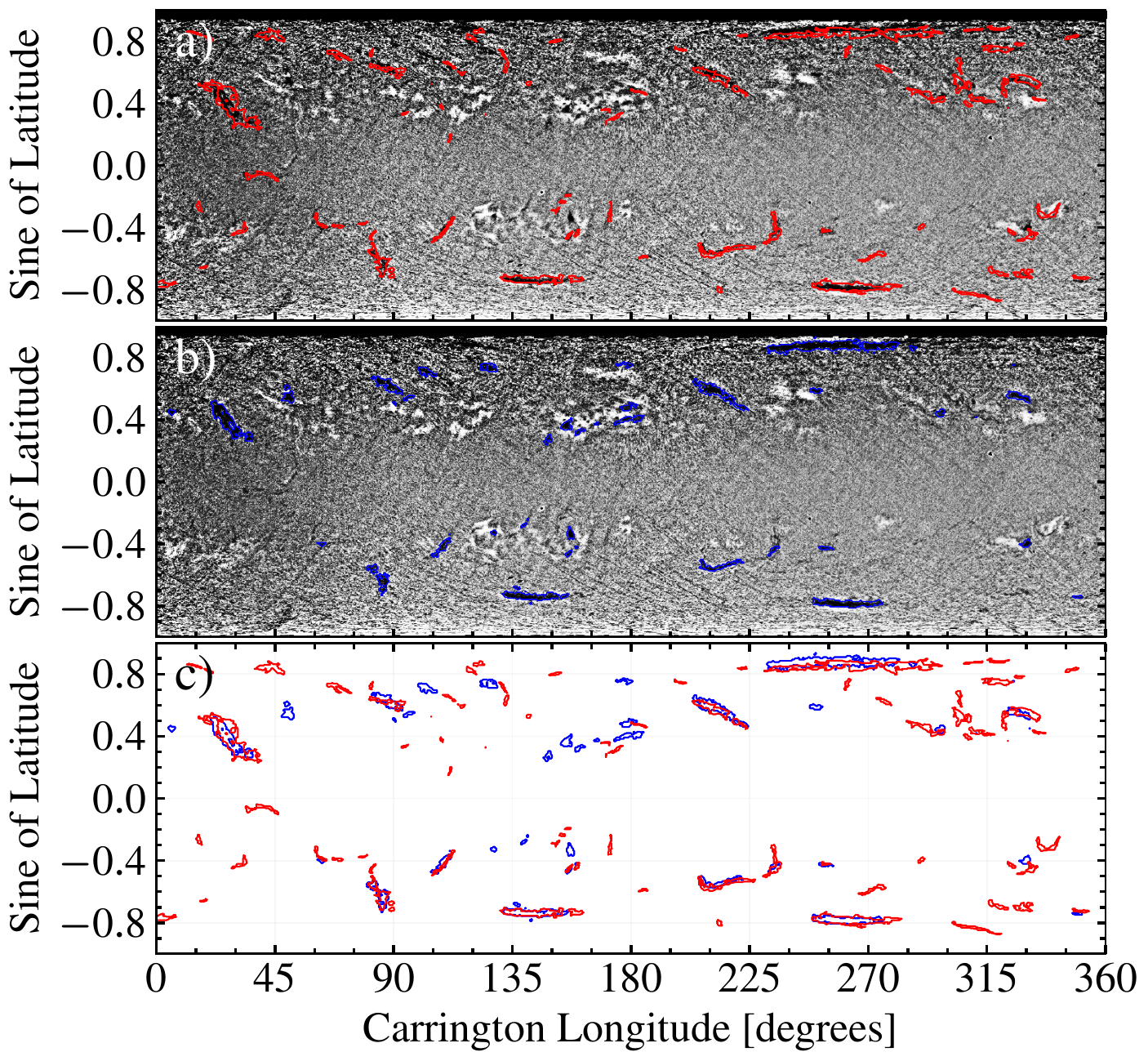}
\caption{(a) and (b) show the Carrington map generated from the H-$\alpha$ plates for CR1371. In (a), the superimposed red contours represent the filaments from the Carrington binary mask generated using suncharts, whereas in (b), the blue contours reflect the filaments from the Carrington binary mask obtained from H-${\alpha}$ plates. The  (c)  shows the overlap of binary masks derived from both sources (suncharts in red and H-$\alpha$ plates in blue). }
\label{fig6}
\end{figure}

In an effort to perform a direct comparison of our results with that of \citet{Chatterjee2017}, who identified filaments using digitized photographic plates of  H-${\alpha}$ observations from Kodaikanal, we also generate Carrington maps using the sunchart data. Figure~\ref{fig6} presents one such comparison. Through these images, we note that many more filaments are detected using the suncharts compared to the H-${\alpha}$ maps. However, we also note instances of over-segmentation in Figure~\ref{fig6}c. In summary, we find a good match between the detections made using the suncharts and H-${\alpha}$ images. Lastly, through figure in Appendix~\ref{add:carr}, we present a different aspect of KoSO suncharts, i.e., a dataset that can also be used as a resource to fill the data gaps in existing catalogues around the globe.

\section{Conclusions} \label{conclu}
In this article, we present, for the first time, the digitised version of the KoSO suncharts that contain multiple solar features such as filaments, plages, sunspots and prominences within one drawing.  Our main findings are the following.

\begin{enumerate}
    \item We devise a novel automatic method to calibrate the sunchart data through disk detection and p-angle correction.
    
    \item We implement the k-means clustering technique to get the optimum threshold, which automatically detects solar filaments from each sunchart over two solar cycles. 

    \item We find a clear `rush to the poles' signature through the time-latitude distribution of filaments. We find a close match of the poleward drift rate, an important parameter to understand the polar field build-up process, with available studies from KoSO digitized H-${\alpha}$ plates for overlapping cycles.
    
    \item Latitudinal distributions of filaments'  length, area and tilt angle from our study show close validation with those from studies utilising other hand-drawn synoptic map archives such as the McIntosh archive and Meudon archive.

\end{enumerate}

Although our automated filament detection procedure works well, for the most part, we have identified some drawbacks too. For example, we had some false detections due to the presence of red patches (likely accidental pen marks) on the images. Moreover, discoloration due to fading sometime results in fragmented filaments. To mitigate these problems, as a future work, we plan to use filament masks produced from this study as ground truth and build a training set to perform supervised machine learning (e.g., convolutional neural networks) to detect filaments. Through this, we can also tackle the cases in which filaments are indicated by green colour (obtained from other observatories).

Our present study corresponding to two solar cycles,  highlights the importance of this independent long-term archive of hand-drawn suncharts. The uniqueness of these suncharts is that they bridge the gap left by damage (fungus, broken plates etc.) on Kodaikanal plate data, allowing us to generate a uniform data series. In near future we expect to provide the full digitised version of the sun charts for almost one century to the solar community.

\section{Acknowledgements}
We thank all the observers who have been involved in observations and making sketches at Kodaikanal for their contributions to building this enormous resource over the last 100 years. We also thank the Department of Science and Technology (DST) for the project grant (DST/ICPS/CLUSTER/Data Science/2018/General/Sl. No.18), which made this digitisation possible. 

\software{IDL, Python, Open-CV, Numpy, Pandas}

\appendix

\section{Filament Detection}
\label{app1}
\begin{figure*}[!ht]
\centering
\includegraphics[width=0.8\textwidth,clip=]{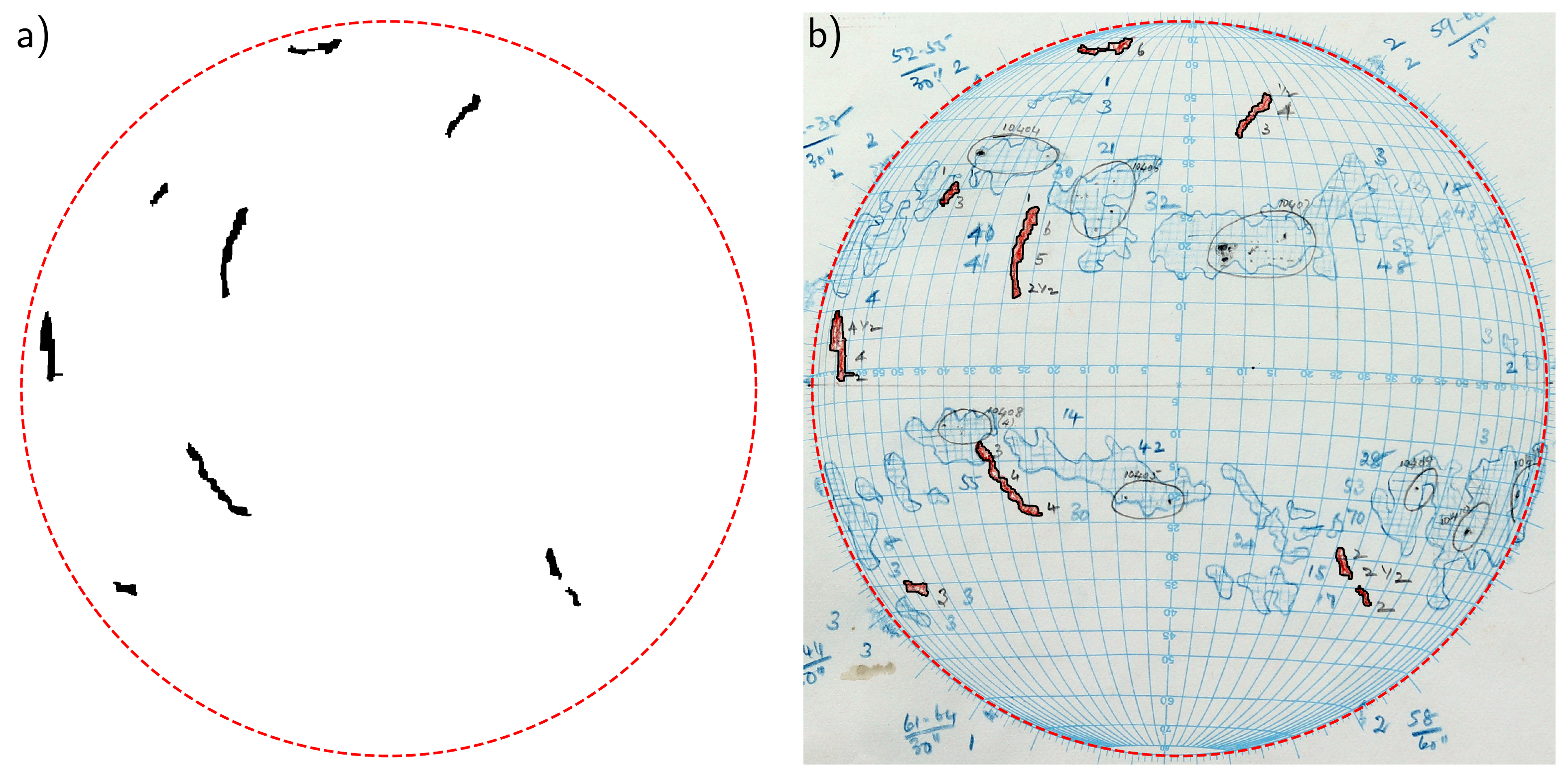}
\caption{Detection of filament with optimal K value. a) A representative image from suncharts with detected filaments, b) Red contours of detected filaments overlaid on original sunchart.}
\label{extra1}
\end{figure*}

\section{Additional Carrington Map}
\label{add:carr}
\begin{figure*}[h]
\centering
\includegraphics[width=\textwidth,clip=]{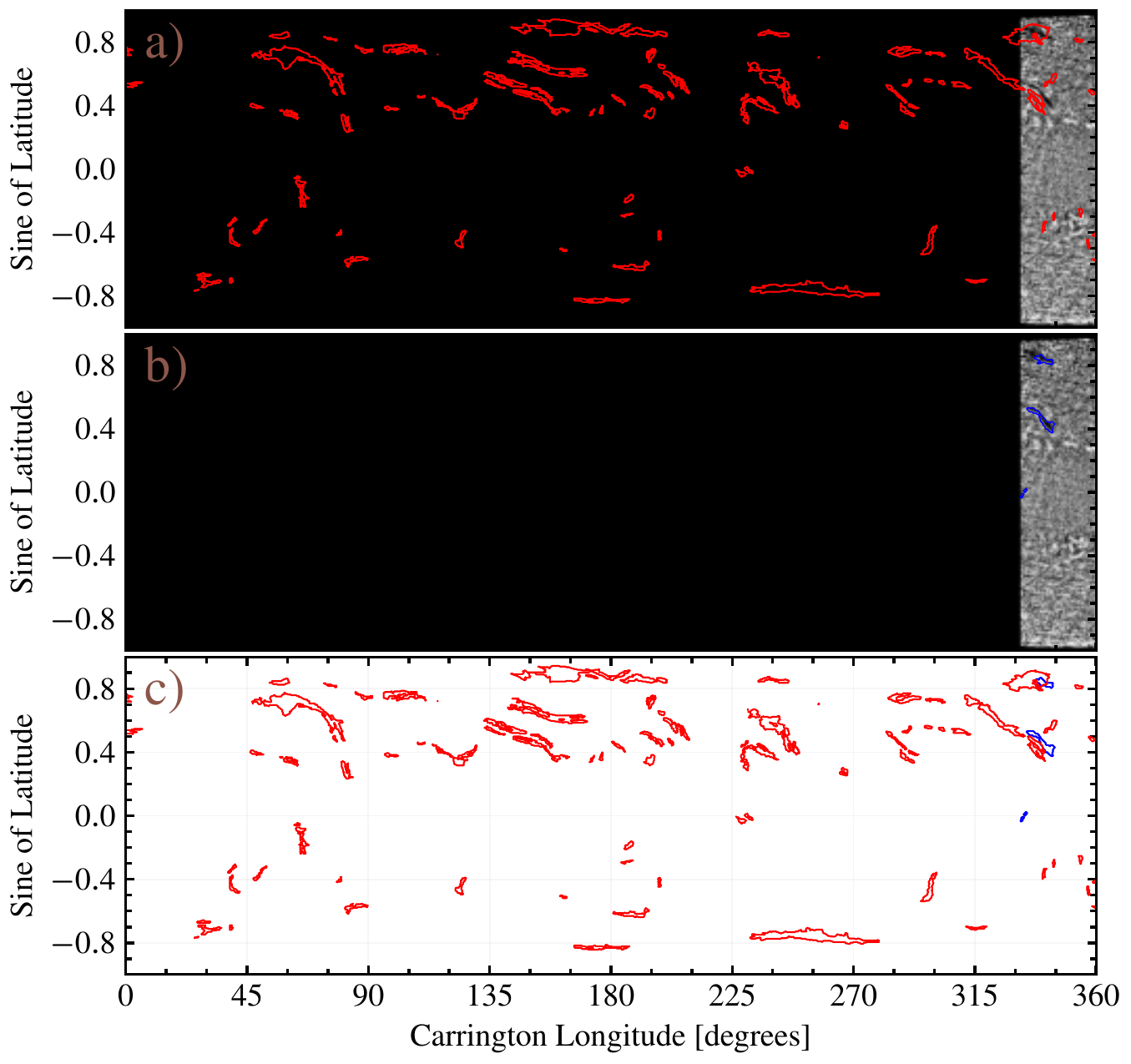}
\caption{Similar as Figure~\ref{fig6} but for the Carrington rotation 1373. In panel b, black patches indicate the missing data. This figure highlights the usefulness of drawings to fill the data gap.}
\label{extra1373}
\end{figure*}

\bibliographystyle{apalike}

\begin{thebibliography}{}

\bibitem[{Ananthakrishnan}, 1952]{Ananthakrishnan1952}
{Ananthakrishnan}, R. (1952).
\newblock {Prominence Activity and the Sunspot Cycle}.
\newblock {\em \nat}, 170(4317):156--158.

\bibitem[{Arlt}, 2008]{Arlt2008}
{Arlt}, R. (2008).
\newblock {Digitization of Sunspot Drawings by Staudacher in 1749 1796}.
\newblock {\em \solphys}, 247(2):399--410.

\bibitem[{Arlt} and {Vaquero}, 2020]{Arlt2020}
{Arlt}, R. and {Vaquero}, J.~M. (2020).
\newblock {Historical sunspot records}.
\newblock {\em Living Reviews in Solar Physics}, 17(1):1.

\bibitem[{Babcock}, 1961]{Babcock1961}
{Babcock}, H.~W. (1961).
\newblock {The Topology of the Sun's Magnetic Field and the 22-YEAR Cycle.}
\newblock {\em \apj}, 133:572.

\bibitem[{Carrasco} et~al., 2020]{Carrasco2020}
{Carrasco}, V.~M.~S., {Gallego}, M.~C., {Arlt}, R., and {Vaquero}, J.~M.
  (2020).
\newblock {On the Use of Naked-eye Sunspot Observations during the Maunder
  Minimum}.
\newblock {\em \apj}, 904(1):60.

\bibitem[{Charbonneau}, 2020]{Charbonneau2020}
{Charbonneau}, P. (2020).
\newblock {Dynamo models of the solar cycle}.
\newblock {\em Living Reviews in Solar Physics}, 17(1):4.

\bibitem[{Chatterjee} et~al., 2017]{Chatterjee2017}
{Chatterjee}, S., {Hegde}, M., {Banerjee}, D., and {Ravindra}, B. (2017).
\newblock {Long-term Study of the Solar Filaments from the Synoptic Maps as
  Derived from ${\rm H}_\alpha$ Spectroheliograms of the Kodaikanal
  Observatory}.
\newblock {\em \apj}, 849(1):44.

\bibitem[Chatterjee et~al., 2020]{chatterjee2020}
Chatterjee, S., Hegde, M., Banerjee, D., Ravindra, B., and McIntosh, S.~W.
  (2020).
\newblock Time-latitude distribution of prominences for 10 solar cycles: A
  study using kodaikanal, meudon, and kanzelhohe data.
\newblock {\em Earth and Space Science}, 7(3):e2019EA000666.
\newblock e2019EA000666 10.1029/2019EA000666.

\bibitem[{Chatzistergos} et~al., 2019]{Theodosios2019d}
{Chatzistergos}, T., {Ermolli}, I., {Solanki}, S.~K., {Krivova}, N.~A.,
  {Giorgi}, F., and {Yeo}, K.~L. (2019).
\newblock {Recovering the unsigned photospheric magnetic field from Ca II K
  observations}.
\newblock {\em \aap}, 626:A114.

\bibitem[{Choudhuri}, 2003]{Arnab2003}
{Choudhuri}, A.~R. (2003).
\newblock {On the Connection Between Mean Field Dynamo Theory and Flux Tubes}.
\newblock {\em \solphys}, 215(1):31--55.

\bibitem[{Fuller} et~al., 2005]{Fuller2005}
{Fuller}, N., {Aboudarham}, J., and {Bentley}, R.~D. (2005).
\newblock {Filament Recognition and Image Cleaning on Meudon
  H{\ensuremath{\alpha}} Spectroheliograms}.
\newblock {\em \solphys}, 227(1):61--73.

\bibitem[{Gao} et~al., 2002]{Gao2002}
{Gao}, J., {Wang}, H., and {Zhou}, M. (2002).
\newblock {Development of an Automatic Filament Disappearance Detection
  System}.
\newblock {\em \solphys}, 205(1):93--103.

\bibitem[{Hao} et~al., 2013]{Hao2013}
{Hao}, Q., {Fang}, C., and {Chen}, P.~F. (2013).
\newblock {Developing an Advanced Automated Method for Solar Filament
  Recognition and Its Scientific Application to a Solar Cycle of MLSO
  H{\ensuremath{\alpha}} Data}.
\newblock {\em \solphys}, 286(2):385--404.

\bibitem[Hough, 1962]{Hough1962}
Hough, P.~V. (1962).
\newblock Method and means for recognizing complex patterns.
\newblock US Patent 3,069,654.

\bibitem[{Jha} et~al., 2022]{Jha2022}
{Jha}, B.~K., {Hegde}, M., {Priyadarshi}, A., {Mandal}, S., {Ravindra}, B., and
  {Banerjee}, D. (2022).
\newblock {Extending the sunspot area series from Kodaikanal Solar
  Observatory}.
\newblock {\em Frontiers in Astronomy and Space Sciences}, 9:1019751.

\bibitem[{Leighton}, 1964]{Leighton1964}
{Leighton}, R.~B. (1964).
\newblock {Transport of Magnetic Fields on the Sun.}
\newblock {\em \apj}, 140:1547.

\bibitem[Lloyd, 1982]{Lloyd1982}
Lloyd, S. (1982).
\newblock Least squares quantization in pcm.
\newblock {\em IEEE Transactions on Information Theory}, 28(2):129--137.

\bibitem[{Low}, 1982]{Low1982}
{Low}, B.~C. (1982).
\newblock {Nonlinear force-free magnetic fields.}
\newblock {\em Reviews of Geophysics and Space Physics}, 20:145--159.

\bibitem[MacQueen, 1967]{MacQueen1967}
MacQueen, J.~B. (1967).
\newblock Some methods for classification and analysis of multivariate
  observations.
\newblock In Cam, L. M.~L. and Neyman, J., editors, {\em Proc. of the fifth
  Berkeley Symposium on Mathematical Statistics and Probability}, volume~1,
  pages 281--297. University of California Press.

\bibitem[{Makarov} and {Sivaraman}, 1983]{Makarov1983}
{Makarov}, V.~I. and {Sivaraman}, K.~R. (1983).
\newblock {Poleward Migration of the Magnetic Neutral Line and the Reversal of
  the Polar Fields on the Sun - Part Two - Period 1904-1940}.
\newblock {\em \solphys}, 85(2):227--233.

\bibitem[{Mandal} et~al., 2017]{Sudip2017}
{Mandal}, S., {Hegde}, M., {Samanta}, T., {Hazra}, G., {Banerjee}, D., and
  {Ravindra}, B. (2017).
\newblock {Kodaikanal digitized white-light data archive (1921-2011): Analysis
  of various solar cycle features}.
\newblock {\em \aap}, 601:A106.

\bibitem[{Mazumder} et~al., 2018]{Mazumder2018}
{Mazumder}, R., {Bhowmik}, P., and {Nandy}, D. (2018).
\newblock {The Association of Filaments, Polarity Inversion Lines, and Coronal
  Hole Properties with the Sunspot Cycle: An Analysis of the McIntosh
  Database}.
\newblock {\em \apj}, 868(1):52.

\bibitem[{Mazumder} et~al., 2021]{Mazumder2021}
{Mazumder}, R., {Chatterjee}, S., {Nandy}, D., and {Banerjee}, D. (2021).
\newblock {Solar Cycle Evolution of Filaments over a Century: Investigations
  with the Meudon and McIntosh Hand-drawn Archives}.
\newblock {\em arXiv e-prints}, page arXiv:2106.04320.

\bibitem[{McIntosh}, 1972]{McIntosh1972}
{McIntosh}, P.~S. (1972).
\newblock {Solar magnetic fields derived from hydrogen alpha filtergrams.}
\newblock {\em Reviews of Geophysics and Space Physics}, 10:837--846.

\bibitem[{Mordvinov} et~al., 2020]{Mordvinov2020}
{Mordvinov}, A.~V., {Karak}, B.~B., {Banerjee}, D., {Chatterjee}, S.,
  {Golubeva}, E.~M., and {Khlystova}, A.~I. (2020).
\newblock {Long-term Evolution of the Sun's Magnetic Field during Cycles 15-19
  Based on Their Proxies from Kodaikanal Solar Observatory}.
\newblock {\em \apjl}, 902(1):L15.

\bibitem[{Pevtsov} et~al., 2016]{Pevtsov2016}
{Pevtsov}, A.~A., {Virtanen}, I., {Mursula}, K., {Tlatov}, A., and {Bertello},
  L. (2016).
\newblock {Reconstructing solar magnetic fields from historical observations.
  I. Renormalized Ca K spectroheliograms and pseudo-magnetograms}.
\newblock {\em \aap}, 585:A40.

\bibitem[{Ravindra} et~al., 2020]{Ravindra2020}
{Ravindra}, B., {Pichamani}, K., {Selvendran}, R., {Samuel}, J., {Kumar}, P.,
  {Jassoria}, N., and {Navneeth}, R.~S. (2020).
\newblock {Sunspot drawings at Kodaikanal Observatory: a representative results
  on hemispheric sunspot numbers and area measurements}.
\newblock {\em \apss}, 365(1):14.

\bibitem[{Senthamizh Pavai} et~al., 2015]{SenthamizhPavai2015}
{Senthamizh Pavai}, V., {Arlt}, R., {Dasi-Espuig}, M., {Krivova}, N.~A., and
  {Solanki}, S.~K. (2015).
\newblock {Sunspot areas and tilt angles for solar cycles 7-10}.
\newblock {\em \aap}, 584:A73.

\bibitem[{Shin} et~al., 2020]{Gyungin2020}
{Shin}, G., {Moon}, Y.-J., {Park}, E., {Jeong}, H., {Lee}, H., and {Bae}, S.-H.
  (2020).
\newblock {Generation of High-resolution Solar Pseudo-magnetograms from Ca II K
  Images by Deep Learning}.
\newblock {\em \apjl}, 895(1):L16.

\bibitem[{Tlatov} et~al., 2016]{Tlatov2016}
{Tlatov}, A.~G., {Kuzanyan}, K.~M., and {Vasil'yeva}, V.~V. (2016).
\newblock {Tilt Angles of Solar Filaments over the Period of 1919 - 2014}.
\newblock {\em \solphys}, 291(4):1115--1127.

\bibitem[{Tlatova} et~al., 2017]{Tlatova2017}
{Tlatova}, K.~A., {Vasil'eva}, V.~V., and {Tlatov}, A.~G. (2017).
\newblock {Reconstruction of a Hundred Years Series of Solar Filaments from
  Daily Observational Data}.
\newblock {\em Geomagnetism and Aeronomy}, 57(7):825--828.

\bibitem[{Usoskin} et~al., 2009]{Usoskin2009}
{Usoskin}, I.~G., {Mursula}, K., {Arlt}, R., and {Kovaltsov}, G.~A. (2009).
\newblock {A Solar Cycle Lost in 1793-1800: Early Sunspot Observations Resolve
  the Old Mystery}.
\newblock {\em \apjl}, 700(2):L154--L157.

\bibitem[{Xu} et~al., 2021]{Xu2021}
{Xu}, Y., {Banerjee}, D., {Chatterjee}, S., {P{\"o}tzi}, W., {Wang}, Z.,
  {Ruan}, X., {Jing}, J., and {Wang}, H. (2021).
\newblock {Migration of Solar Polar Crown Filaments in the Past 100 Years}.
\newblock {\em \apj}, 909(1):86.

\bibitem[{Yuan} et~al., 2011]{Yuan2011}
{Yuan}, Y., {Shih}, F.~Y., {Jing}, J., {Wang}, H., and {Chae}, J. (2011).
\newblock {Automatic Solar Filament Segmentation and Characterization}.
\newblock {\em \solphys}, 272(1):101--117.

\bibitem[{Zhu} et~al., 2019]{Zhu2019}
{Zhu}, G., {Lin}, G., {Wang}, D., {Liu}, S., and {Yang}, X. (2019).
\newblock {Solar Filament Recognition Based on Deep Learning}.
\newblock {\em \solphys}, 294(9):117.

\end{thebibliography}

\end{document}